\documentclass[%
 reprint,
superscriptaddress,
nofootinbib,
 amsmath,amssymb,
 aps,
 prl,
floatfix,
]{revtex4-2}

\usepackage{enumitem}

\usepackage{graphicx}
\usepackage{dcolumn}
\usepackage{bm}

\usepackage{epsfig}
\usepackage{amsmath}
\input{epsf}
\usepackage{psfrag}
\usepackage{xcolor}
\usepackage{pdfpages}
\usepackage[normalem]{ulem}
\makeatletter
\AtBeginDocument{\let\LS@rot\@undefined}
\makeatother
\newcommand{\bea}{\begin{eqnarray}}
\newcommand{\eea}{\end{eqnarray}}

\newcommand{\nn}{\nonumber}

\newcommand{\ket}[1]{\left| #1\right>}

\def\tS{{\tt S}}
\def\tT{{\tt U}}
\def\tU{{\tt T}}

\begin{document}

\title{Long Range Energy-energy Correlator at the LHC}

\author{Yuxun Guo}%
 \email{yuxunguo@lbl.gov}
\affiliation{Nuclear Science Division, Lawrence Berkeley National Laboratory, Berkeley, CA 94720, USA}%

\author{Xiaohui Liu}
 \email{xiliu@bnu.edu.cn}
 \affiliation{Center of Advanced Quantum Studies, Department of Physics, Beijing Normal University, Beijing, 100875, China}
 \affiliation{Key Laboratory of Multi-scale Spin Physics, Ministry of Education, Beijing Normal University, Beijing 100875, China}
 
\author{Feng Yuan}%
 \email{fyuan@lbl.gov}
\affiliation{Nuclear Science Division, Lawrence Berkeley National Laboratory, Berkeley, CA 94720, USA}%
\affiliation{Institute for Theoretical Physics,
                Universit\"{a}t T\"{u}bingen,
                Auf der Morgenstelle 14,
                D-72076 T\"{u}bingen, Germany}

\begin{abstract}
We study the forward-backward azimuthal angular correlations of hadrons in association with multi-particle production in the central rapidity region in proton-proton collisions at the LHC. We apply the nucleon energy-energy correlator framework, where the spinning gluon distribution introduces a nontrivial $\cos(2\phi)$ asymmetries. We will demonstrate that the fundamental helicity structure of QCD amplitudes predicts a unique power counting rule: $\cos(2\phi)$ asymmetry starts at ${\cal O}(\alpha_s^2)$ order for dijet, ${\cal O}(\alpha_s)$ for three jet and ${\cal O}(1)$ for four (and more) jet productions. Our results will help us to understand the long standing puzzle of nearside ridge behavior observed in high multiplicity events of $pp$ collisions at the LHC.   
\end{abstract}

\maketitle

\textbf{\textit{  Introduction.}} More than a decade ago, a long range azimuthal angular correlation of $\cos(2\phi)$ between two hadrons, the so-called nearside ridge, was discovered in high multiplicity events in proton-proton ($pp$) collisions at the LHC, where the event generator, such as PYTHIA, has shown very small (close to zero) asymmetry~\cite{CMS:2010ifv}. This has stimulated tremendous experimental and theoretical efforts in the last decade, but the puzzle still stands~\cite{CMS:2015fgy,CMS:2016fnw,ATLAS:2015hzw,ATLAS:2016yzd,ALICE:2021nir,ALICE:2023ulm,Dusling:2012iga,Dusling:2013oia,Dumitru:2010iy,Bzdak:2013zma,Dusling:2015gta,Loizides:2016tew,Strickland:2018exs,Nagle:2018nvi,Baty:2021ugw,CMS:2023iam,Grosse-Oetringhaus:2024bwr}. In this paper, we investigate the long range correlation from different perspective, applying the nucleon energy-energy correlator (NEEC)~\cite{Liu:2022wop,Cao:2023oef,Li:2023gkh} at the LHC. We will show that the spinning gluon distribution in this framework~\cite{Li:2023gkh} leads to a nonzero $\cos(2\phi)$ azimuthal asymmetries in forward-backward energy correlators in $pp$ collisions and a power counting rule predicts a strong coupling constant $\alpha_s$ hierarchy depending on the number of jets in the central rapidity region . Experimental verification of this observation will help us to understand the nearside ridge behavior in $pp$ collisions. 

Previous studies of NEECs focus mainly on the deep inelastic scattering (DIS)~\cite{Liu:2022wop,Cao:2023oef,Liu:2023aqb,Li:2023gkh} at the planned Electron-Ion Collider (EIC)~\cite{Accardi:2012qut,AbdulKhalek:2021gbh,Proceedings:2020eah}. In Ref.~\cite{Guo:2024jch}, we apply the concept of NEEC to the observable in $pp$ collisions at the LHC. Here, one can measure the energy-energy correlators along the beam directions of both incoming hadrons with polar angles $\theta_{a,b}$ and azimuthal angles $\phi_{a,b}$, respectively. The hard scattering, such as jet production studied in this paper, happens in the central rapidity region. For small polar angles of $\theta_{a}$ and $\theta_b$ in opposite directions, their rapidity difference will be large, which corresponds to a long range correlation. From the results presented in Ref.~\cite{Guo:2024jch}, we find that the asymmetries in Higgs Boson and top quark pair productions are quite sizable with opposite signs. The goal of the current paper is to carry out multi-jet production.

\begin{figure}[htbp]
  \begin{center}
   \includegraphics[width=0.48\textwidth]{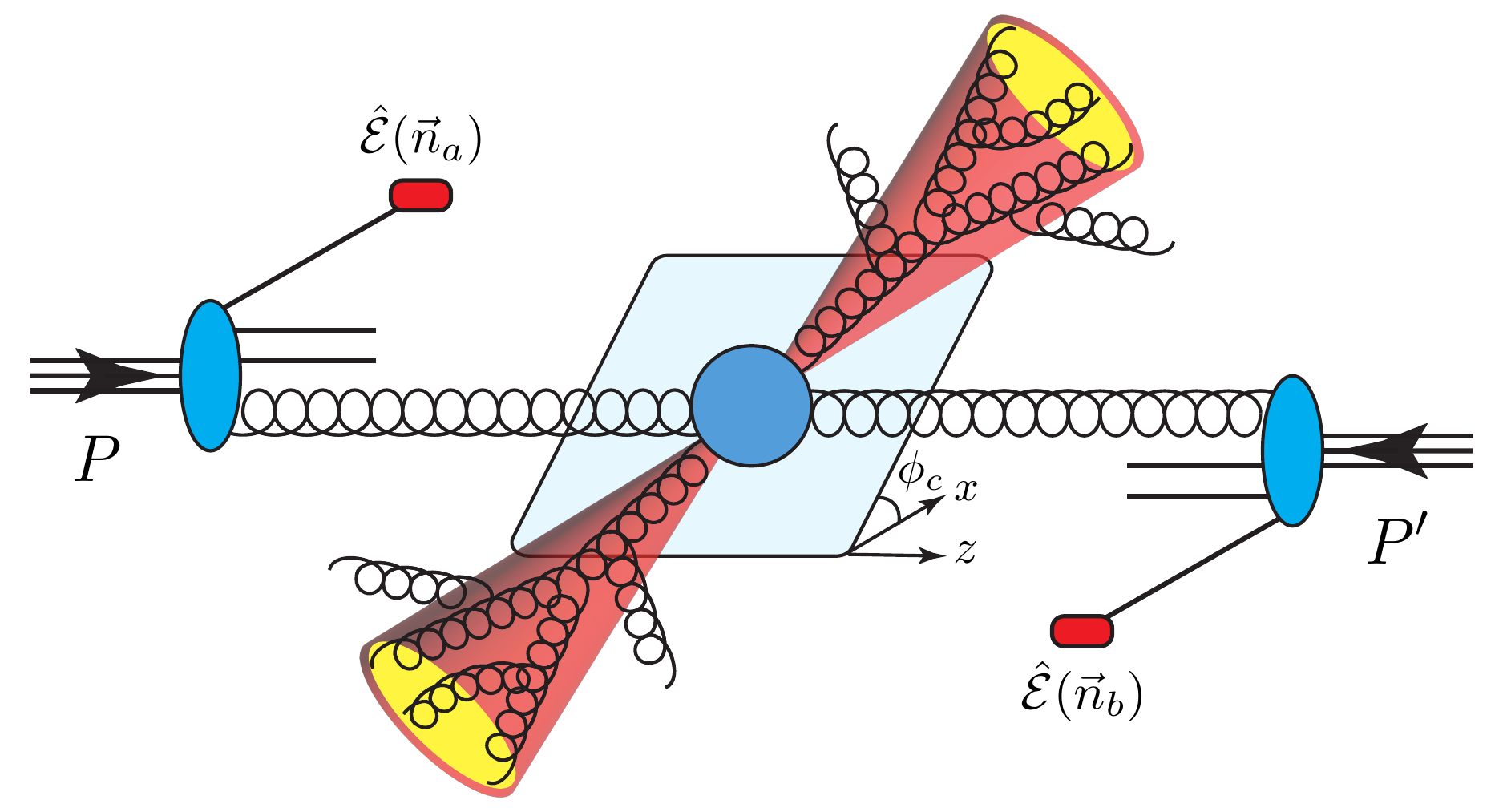} 
\caption{Forward-backward energy-energy correlators in proton-proton collisions at the LHC where multi-jets are produced in the central rapidity region. The spinning gluons will lead to a long range $\cos(2\phi_a-2\phi_b)$ angular correlation. 
}
  \label{fg:measure}
 \end{center}
  \vspace{-5.ex}
\end{figure}

We will show that the long range $\cos(2\phi)$ azimuthal correlation is closely related to the fundamental structure of QCD helicity amplitudes~\cite{Parke:1986gb,Berends:1987me}. In addition, jet production plays an important role in high multiplicity events in high energy hadronic collisions, see, for example, Ref.~\cite{Mueller:1981ex}.
In $pp$ collisions at the LHC, these events are normally selected by the number of particles detected in the center~\cite{CMS:2010ifv}. Therefore, our study in the following will lead a new way to decipher the underlying mechanism of the near side ridge. 

For jet production at the LHC, the gluon NEEC is the dominant contribution and can be defined as~\cite{Li:2023gkh,Guo:2024jch},
\bea\label{eq:fgmunu}
&& f^{\alpha\beta}_{g,{\rm EEC}} (x,\vec{n}_a)
= 
\int \frac{dy^-}{2\pi xP^+ } e^{- i x P^+ \frac{y^-}{2} }   \nn \\ 
&&
\hspace{4.ex} 
\times  
\langle P| 
{\cal F}^{+\alpha}
\left(y^- \right) 
{\cal L}^\dagger[\bm{\infty},y^-]
\hat{{\cal E}}({\vec n}_a)   
{\cal L} [\bm{\infty},0]
{\cal F}^{+\beta}
(0)  |P \rangle  \,
\nn \\ 
&&
= \left(-{g_T^{\alpha\beta}}/{2}\right) f_{g,{\rm EEC}}
+ h_T^{\alpha\beta}
d_{g,{\rm EEC}}\,,
\eea 
where $F^{\mu\nu}$ represents the gluonic field strength tensor. Because of the spin-$1$ feature, we have two independent gluon NEECs: the diagonal one $f_{g,{\rm EEC}}$ projected from $g_T^{\alpha\beta} = g^{\alpha\beta} - (P^\alpha{\bar n}^\beta +  {\bar n}^\alpha P^\beta)/{{\bar n}\cdot P}$ corresponds to the usual gluon distribution, whereas the off-diagonal one $d_{g,{\rm EEC}}$ from $h_T^{\alpha\beta}=n_{a,T}^\alpha n_{a,T}^\beta/|n_{a,T}^2|+g_T^{\alpha\beta}/2$ represents the spinning gluon. Here ${\bar n}\cdot P = P^0+P^z \equiv P^+$ and $n_{a,T}^\alpha = (0,\vec{n}_a,0)$ is the transverse component of $n_{a}^\alpha$. The above definition is for the gluon NEEC associated with the proton moving in $+\hat z$ direction with momentum $P$. A similar definition can be written for the gluon NEECs for the proton moving in $-\hat z$ direction with momentum ${P}'$ and energy flow direction $n_b^\alpha = (1,\sin\theta_b \cos\phi_b, \sin\theta_b \sin \phi_b, \cos\theta_b)$. 

\begin{figure}[htbp]
  \begin{center}
   \includegraphics[scale=0.5]{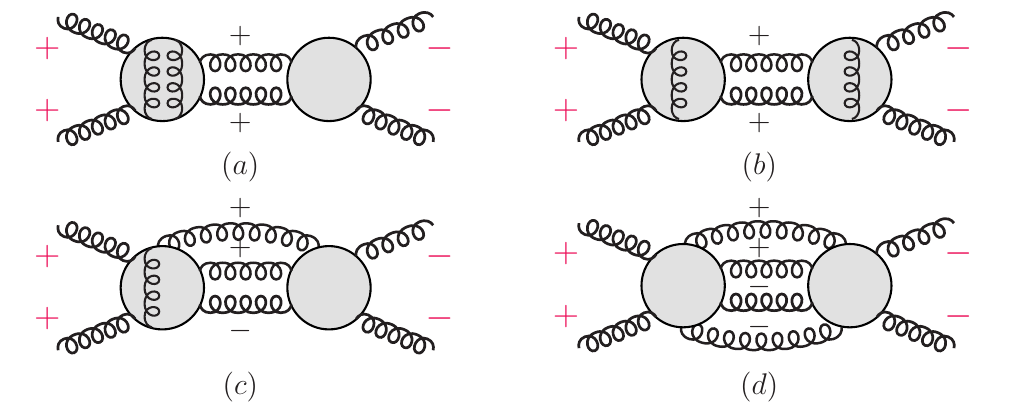} 
\caption{Nonzero $\cos(2\phi)$ asymmetry comes from the interference between double helicity-flip amplitudes with positive helicity for both incoming gluons. 
Example of helicity amplitudes that contribute to  $\cos(2\phi)$ asymmetries in dijet (order $\alpha_s^2$) from $(a,b)$, three jets (order $\alpha_s$) from $(c)$ and four jets (order 1) from $(d)$ production in $pp$ collisions. Cuts through the intermediate states are implied. }
  \label{fg:feyn}
 \end{center}
  \vspace{-5.ex}
\end{figure}

The spinning gluon originates from the interference between different helicity states. 
To generate a long range correlation between $\vec{n}_a$ and $\vec{n}_b$, we need to couple two $d_{g,{\rm EEC}}(\theta)$ from both incoming protons. In addition, the hard scattering processes in the mid-rapidity act as a polarizer that leads to different sizes and signs of the $\cos(2\phi)$ asymmetry, where $\phi=\phi_a-\phi_b$. For example, the Higgs Boson production produces sizable positive $\cos(2\phi)$ asymmetry while the top quark pair results in a negative asymmetry~\cite{Guo:2024jch}. For multi-jet production, as we will show below, the $\cos(2\phi)$ asymmetry depends on the helicity structure of partonic amplitudes. As a result, it vanishes at the leading order (LO) and next-to-leading order (NLO) for dijet production. We will derive the explicit results for dijet production at ${\cal O}(\alpha_s^2)$ order. Meanwhile, $\cos(2\phi)$ asymmetry starts at ${\cal O}(\alpha_s)$ for three jet and ${\cal O}(1)$ for four (and more) jet productions, as illustrated in Fig.~\ref{fg:feyn}.

\textbf{\textit{Azimuthal $\cos(2\phi)$ Asymmetry for Dijet Production. }} 
We start with two particle production in the mid-rapidity. The generic cross section measurement for forward-backward correlation takes the following form,
\bea\label{eq:e3c} 
 \Sigma^{jet}(Q^2; \theta_{a,b},\phi)&=&\sum_{ij}\int  d\sigma^{jet}(Q^2)\frac{E_i}{E_P} \frac{E_j}{E_P} {\cal F}(\phi; \vec{n}_{a,b})  \nn \\ 
&& ~~~~~~\times
\delta(\vec{n}_a-\vec{n}_i)\delta(\vec{n}_b-\vec{n}_j) \nn\\
&=&\int d\Omega \left[\sigma_0+\sigma_2\cos(2\phi)\right] \,,
\eea 
where $Q$ represents the hard momentum scale such as the jet energy, ${\cal F}(\phi; \vec{n}_{a,b})$ imposes the phase space measurement to construct $\phi$ and $\int d\Omega$ for the part related to multi-jet production in the central rapidity. 
The factorization formula for $\sigma_0$ and $\sigma_2$ can be written as,
\bea\label{eq:sig-z} 
\sigma_0&=& x_af_{g,{\rm EEC}}\left(x_a,\theta_a^2  \right)x_bf_{g,{\rm EEC}}\left(x_b,\theta_b^2  \right) \hat{\sigma}_0\\
\sigma_2&=&
x_ad_{g,{\rm EEC}}\left(x_a,\theta_a^2  \right)x_bd_{g,{\rm EEC}}\left(x_b,\theta_b^2  \right)\hat\sigma_2 
\ , 
\eea 
where $\hat\sigma_{0,2}$ represents the partonic cross sections for multi-jet productions and for simplicity we omit the superscript ``$jet$". The final results of $\cos(2\phi)$ asymmetries depend on the NEEC gluon distributions, for which we will follow Ref.~\cite{Guo:2024jch} and take the leading power contributions. 

Same as those in Higgs Boson and top quark pair productions~\cite{Guo:2024jch}, $\cos(2\phi)$ asymmetries come from interference between the double helicity flip amplitudes with positive helicities for both incoming gluons, see Fig.~\ref{fg:feyn}. This leads to a unique prediction as a result of the helicity structure of QCD amplitudes~\cite{Parke:1986gb,Berends:1987me}. Let us start with dijet production in $gg\to gg$ channel. The $\cos(2\phi)$ comes from the interference between the following amplitudes,
\begin{eqnarray}
    \hat \sigma_2&\propto & \sum_{\lambda_3\lambda_4}
    {\cal A}(\pm,\pm,\lambda_3,\lambda_4){\cal A}^*(\mp,\mp,\lambda_3,\lambda_4) \ ,\label{eq:sigma2gg}
\end{eqnarray}
where the two incoming gluons have the same helicity but they are flipped in the complex conjugate of the amplitude. 
It is well known that the tree-level amplitudes vanish where either all the external gluons have the same helicity or only one of them has a different helicity~\cite{Parke:1986gb,Berends:1987me}. Applying this rule to the $gg\to gg$ amplitudes, we have ${\cal A}^{(0)}(++++)={\cal A}^{(0)}(+++-)=0$. As a result, $\hat\sigma_2$ vanishes for any combination of $\lambda_3$ and $\lambda_4$ in Eq.~(\ref{eq:sigma2gg}).
At one-loop order, although the above two amplitudes no longer vanish, there is no contribution from their interference with ${\cal A}^{(0)}$ due to orthogonal color structures. For example, non-vanishing $A^{(1)}(++++)$ has the following color structure~\cite{Bern:1991aq,Kunszt:1993sd,Bern:2002tk,Caola:2021izf}, 
$A^{(1)}(++++)
\propto
\left({\rm Tr}\left[T^aT^bT^cT^d\right]+{\rm Tr}\left[T^aT^cT^dT^b\right]
+{\rm Tr}\left[T^aT^dT^bT^c\right]\right)$, 
where $a$, $b$, $c$, $d$ are color indices for the incoming two gluons and outgoing two gluons, respectively.
If we couple this with the color structure of the leading order amplitude which can be written as
$A_s^{(0)}f_{abe}f_{cde}+A_t^{(0)}f_{ace}f_{bde}+A_u^{(0)}f_{ade}f_{bce}$, 
it does not yield any results. That means the $\cos(2\phi)$ vanishes at $\alpha_s$ order for dijet production too.
Therefore, nonzero $\hat\sigma_2$ for dijet production only starts at NNLO.

At this order, we will have the following helicity amplitude contributions: 
$A^{(1)}(+++-)A^{(1)*}(--+-) $, $A^{(1)}(++++)A^{(1)*}(--++)$, and $A^{(2)}(++++)A^{(0)*}(--++)$.
The first term is easy to derive because both amplitudes come from non-vanishing leading contribution at ${\cal O}(\alpha_s)$.
On the other hand, the derivation of the second and third terms at NNLO is much more involved, due to nontrivial features in these amplitudes that lead to an emerging infrared (IR) divergence even though they vanish at NLO. 
For example, we can apply the NLO amplitudes derived in Refs.~\cite{Bern:1991aq,Kunszt:1993sd,Bern:2002tk,Ahmed:2019qtg,Caola:2021izf}, expressed in dimensional regulation with $D=4-2\epsilon$, and find that the NLO contribution $\hat\sigma_2^{(1)}$ is proportional to $\epsilon$, i.e., $\hat{\sigma}_2^{(1)}=\epsilon \sigma_2^{(1)\epsilon}$, where
\begin{equation}
       \epsilon \sigma_2^{(1)\epsilon}= \frac{2}{{\cal V}}\left<A^{(1)}(++++)|A^{(0)}(--++)\right> + h.c.
    \ ,\label{eq:s10e}
\end{equation}
where ${\cal V}=4(N^2-1)^2$ represents the color and spin averaging factor. 
Here and in the following, we expand the perturbative corrections in terms of $(\alpha_s/4\pi)^i$. However, the soft gluon radiation at NNLO will generate terms proportional to $1/\epsilon^2$, resulting into an $1/\epsilon$ divergence from both real and virtual diagrams. Additional divergence comes from anomalous soft gluon radiation due to color space modification. 
These divergences will be canceled out between real and virtual contributions. 

In particular, the singular behavior of QCD amplitudes has been well established at two loop order~\cite{Catani:1998bh} and can be further factorized into color-independent jet function and color-correlated soft function~\cite{Sterman:2002qn}, 
\begin{equation}
    \boldsymbol{I}^{(1)}\equiv \Bigg[-\sum_i\left(\frac{\gamma_K^{[i](1)}}{2\epsilon^2}+\frac{\mathcal{G}_0^{[i](1)}}{\epsilon}\right)\boldsymbol{1}+\frac{\boldsymbol{\Gamma}^{(1)}}{\epsilon}\Bigg]\left(-\frac{\mu^2}{s}\right)^\epsilon \ .
\end{equation}
The first term associated with the identity matrix $\boldsymbol{I}$ is color-uncorrelated. It originates from the jet function of each parton in the process and thus sums over all partons $i$. The second term is collected by the so-called soft anomalous dimension matrix $\boldsymbol{\Gamma}^{(1)}$ which is color-correlated. It depends on the process, and can be generally written as~\cite{Aybat:2006wq},
\begin{equation}
    \boldsymbol{\Gamma}^{(1)}=\frac{1}{2}\sum_i \sum_{j\not = i} \boldsymbol{T}_i \cdot \boldsymbol{T}_j \ln\left(\frac{-\mu^2}{s_{ij}}\right)\ ,
\end{equation}
where $s_{ij}=(p_i+p_j)^2$ in all outgoing notation. 
Applying the above procedure, the IR divergence of NNLO $\hat\sigma_2^{(2)}$ can be written as,
\begin{equation}
    \hat\sigma_2^{(2)v} = \frac{1}{\epsilon}\left( \hat{\sigma}_{2,J}^{(2)v} + \hat{\sigma}_{2,S}^{(2)v} \right) + \rm{finite~terms} \ ,
\end{equation}
where
\begin{eqnarray}
    \hat{\sigma}_{2,S}^{(2)v}&= &\frac{2}{{\cal V}}\left<A^{(1)}(++++)|2\text{Re}\left[\boldsymbol{\Gamma}^{(1)}\right]|A^{(0)}(--++)\right>\nn\\
    &&+ h.c. \ ,\nonumber \\
    \hat{\sigma}_{2,J}^{(2)v}&= &-\sum_i  \gamma_K^{[i](1)} \hat{\sigma}_2^{(1)\epsilon} = - 8 C_A \hat{\sigma}_2^{(1)\epsilon} \ , \label{eq:softvirtual}
\end{eqnarray}
with $\gamma_K^{[g](1)}=2C_A$.

The real gluon radiation can be derived in a similar manner, following previous examples~\cite{Botts:1989kf,Kidonakis:1997gm,Kidonakis:1998bk,deFlorian:2013qia,Sun:2014gfa,Sun:2015doa}. In particular, their contributions are formulated in the same color space as a soft radiation matrix $\overline{\cal S\!\!S}$, 
\begin{equation}
    \hat \sigma_{2,S}^{(2)r} =  \left<[\cdots]|\overline{\cal S\!\!S}|[\cdots]\right> + h.c.\ ,
\end{equation}
where the $[\cdots]$ stands for the helicity amplitude in the color space. The only difference is that the real gluon contributes to a finite transverse momentum $q_\perp$, which is taken much smaller than the dijet transverse momentum $P_\perp$ to differentiate from three jets in the final state. The leading contributions are obtained by applying Eikonal approximation for all soft gluons attached to the external gluons and  
can be equivalently expressed in terms of the following matrix $\overline{\cal S\!\!S}$,
\begin{eqnarray}
     \overline{{\cal S\!\!S}}&=& 
        4\pi\left(\frac{1}{ q_\perp^2}2\text{Re}[\boldsymbol{\Gamma}^{(1)}] +4N\frac{1}{q_\perp^2} \ln\frac{\hat s }{q_\perp^2 R^2}\boldsymbol{1}\right)+\cdots\ ,
    \label{eq:softmatrix}
\end{eqnarray} 
where $R$ is the jet size and $\cdots$ stands for terms that contribute to finite terms of $\hat \sigma_{2,S}^{(2)r}$. We have applied a narrow jet approximation to derive the $R$-dependence in the above soft gluon radiation~\cite{Mukherjee:2012uz,Sun:2014gfa}. The integral over $q_\perp$ will lead to IR divergences,
\begin{eqnarray}
&\!\!&\!\!
\mu^{2\epsilon}
\int^{q_0}\frac{d^{d}q_\perp}{(2\pi)^{d}}\frac{1}{q_\perp^2}=\frac{1}{4\pi}\left(-\frac{1}{\epsilon}+\ln\frac{q_0^2}{\mu^2}\right )\ ,\\
&\!\!&\!\!
\mu^{2\epsilon}
\int^{q_0}\frac{d^{d}q_\perp}{(2\pi)^{d}}\frac{1}{q_\perp^2}\ln\frac{\hat s}{q_\perp^2}=\frac{1}{4\pi}\left(\frac{1}{\epsilon^2}-\frac{1}{\epsilon}\ln\frac{\hat s}{\mu^2}+\cdots\right)\ ,
\end{eqnarray}
where $d=2-2\epsilon$ and we have introduced an upper bound $q_0$ in $q_\perp$, which is interpreted as the jet resolution parameter. Additional IR divergences come from jet functions from the two gluon jets~\cite{Mukherjee:2012uz,Sun:2014gfa}, 
\begin{equation}
    2{\cal J}_g=2\frac{\alpha_s N}{2\pi}\left[\frac{1}{\epsilon^2}+\frac{1}{\epsilon}\left(2\beta_0-\ln\frac{P_\perp^2R^2}{\mu^2}\right)+\cdots\right]\ ,  \label{eq:jet}  
\end{equation}
which is also multiplied by $\epsilon\hat\sigma_{2}^{(1)\epsilon}$. With all these results, we show that the IR divergences are canceled out completely, in particular, $\hat\sigma_{2,S}^{(2)v}$ is canceled by the first term of Eq.~(\ref{eq:softmatrix}), $\hat\sigma_{2,J}^{(2)v}$ by the second term of (\ref{eq:softmatrix}) and the jet function (\ref{eq:jet}). 
After canceling all IR divergences and jet size $R$ dependence, we arrive at the following result,
\begin{eqnarray}
\sigma_2&=&\int\frac{d\xi}{\xi}\frac{d\xi'}{\xi'}x_ad_{g,\rm EEC}(x_a')x_bd_{g,\rm EEC}(x_b')\left\{-\hat\sigma_{2,1}^{(2)}(x)\right.\nn\\
&\times &\left[{\cal P}_{gg}^d(\xi)\delta(1-\xi')+{\cal P}_{gg}^d(\xi')\delta(1-\xi)\right]\label{eq:dijetfinal}\\
&+&\left.\delta(1-\xi)\delta(1-\xi')\left[\hat\sigma_{2,0}^{(2)}(x)\ln\left(
 \frac{\hat s}{q_0^2}\right)+\hat\sigma_{2,2}^{(2)}(x)\right]\right\} \ , \nn 
 \end{eqnarray}
where $x=-t/s$, $x_{a}'=x_{a}/\xi$, $x_{b}'=x_{b}/\xi'$, ${\cal P}_{gg}^d(\xi)=\xi/(1-\xi)_++\beta_0\delta(1-\xi)$ with $\beta_0=11/12-N_f/18$. In the above, $\hat\sigma_{2,i}^{(2)}$ depends on $x$, 
\begin{eqnarray}
    \hat\sigma_{2,0}^{(2)}&=&-
(N^2+6)(N^2-N_f)\left(\frac{\ln x}{\bar x}+\frac{\ln\bar x}{x}\right)\ ,\\
\hat\sigma_{2,1}^{(2)}&=&\frac{N^2(N_f-N)}{2}\left(
\pi^2-2\ln\bar x\ln x+\frac{1+x^2}{x^2}\ln^2{\bar x}\right.\nonumber\\
&&\left. +\frac{1+\bar x^2}{\bar x^2}\ln^2{ x}+ \frac{4\ln\bar x}{x}+\frac{4\ln x}{\bar x}
\right)\ ,\\
\hat\sigma_{2,2}^{(2)}&=&\frac{1}{24(x\bar x)^2}\left\{
12(N-N_f)^2(1-x\bar x)^2(2-x\bar x)\right.\nn\\
&&+5(N^2-N_f)x\bar x\left[2(N^2-N_f)x\bar x(x-\bar x)\ln\frac{x}{\bar x}\right.\nn\\
&&\left.\left.-\omega_1(x)\left(\ln^2\frac{\bar x}{x}+\pi^2\right)
-2N^2\omega_2(x)
\right]\right\}\nn\\
&&-15(N^2-N_f)\frac{\ln x\ln\bar x}{x\bar x}\ ,
\end{eqnarray}
where $\bar x=1-x$, $\omega_1(x)=2N^2(1+2x\bar x-x^2\bar x^2)-N_f(x\bar x -2x^2\bar x^2)$ and $\omega_2(x)=4\ln x\ln\bar x-x\ln^2\bar x-\bar x\ln^2 x$.

\begin{figure}[htbp]
  \begin{center}
   \includegraphics[scale=1]{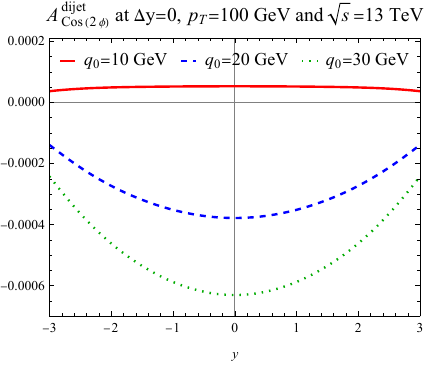} 
\caption{Long range $\cos(2\phi)$ azimuthal asymmetries in nucleon EEC observable associated with dijet production as a function of the rapidity $y$ of the two jets.}
  \label{fg:dijet2phi}
 \end{center}
  \vspace{-5.ex}
\end{figure}

In Fig.~\ref{fg:dijet2phi}, we show the numerical results for the $\cos(2\phi)$ asymmetries as functions of the rapidity $y$ for the dijet for the kinematics at the LHC with $P_\perp=100\rm GeV$. The contribution from the real soft gluon radiation crucially depends on the upper bound $q_0$, which should be much smaller than the jet transverse momentum $P_\perp$. To illustrate this dependence, three values of $q_0$ have been chosen in Fig.~\ref{fg:dijet2phi}.  can be regarded as a jet veto. Because our derivation only applies to soft gluon, we restrict the kinematic region of $q_0\ll P_\perp$. Another interesting feature of Eq.~(\ref{eq:dijetfinal}) is that there is a cancellation between the two dominant terms (the first two) because they have opposite signs. For some particular choice of $q_0$, they cancel out each other completely. The experiment observation of this behavior will provide an important confirmation of our predictions. 

\begin{figure}[htbp]
  \begin{center}
   \includegraphics[scale=0.95]{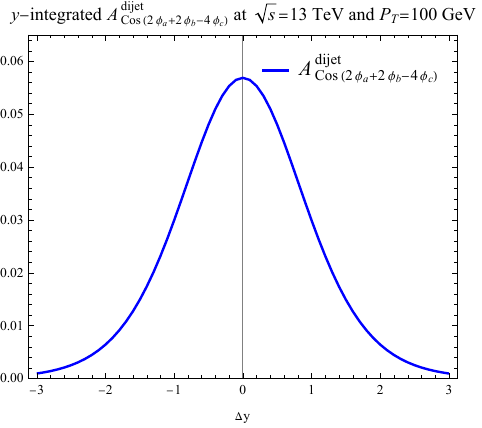} 
\caption{$\cos(2\phi_a+2\phi_b-4\phi_c)$ azimuthal asymmetries associated with dijet production as a function of the rapidity difference between the two jets $\Delta y$. }
  \label{fg:dijet}
 \end{center}
  \vspace{-5.ex}
\end{figure}

The spinning gluon $d_{g,\rm EEC}(x_a)$ can also contribute to azimuthal asymmetries which depend on the dijet production plane. This can be utilized as a cross check of our derivations. In Fig.~\ref{fg:dijet}, we show the azimuthal correlations between $\phi_a$, $\phi_b$ and $\phi_c$ for the dijet production, where $\phi_c$ is the azimuthal angle for the jet transverse momentum. From our calculations, we find that the $\cos(2\phi_a+2\phi_b-4\phi_c)$ asymmetry is nonzero,
\begin{eqnarray}
    &&\langle \cos(2\phi_a+2\phi_b-4\phi_c)\rangle_{\rm dijet}\nonumber\\
    &&~~~~=\frac{\int d\Omega d_{g,\rm EEC}(x_a)d_{g,\rm EEC}(x_b) \frac{9}{2}\frac{\hat s^2+\hat t^2+\hat u^2}{2\hat s^2}}{\int d\Omega f_{g,\rm EEC}(x_a)f_{g,\rm EEC}(x_b)\frac{9}{2}\frac{(\hat s^2+\hat t^2+\hat u^2)^3}{8\hat s^2\hat t^2\hat u^2}} \ .
\end{eqnarray}
This asymmetry also comes from interference between double helicity flip amplitudes but the incoming gluons have opposite helicities, i.e., ${\cal A}^{(0)}(+-+-){\cal A}^{(0)*}(-++-)$ which is nonzero at the leading order. Numerically, $\langle \cos(2\phi_a+2\phi_b-4\phi_c)\rangle$ is about $10\%$ for the typical kinematics at the LHC. It will vanish when we average over $\phi_c$ and will not contribute to the long range $\cos(2\phi)$ asymmetries discussed in previous examples. However, they come from the same physics, and it is important to cross check this prediction.

\textbf{\textit{Discussion and Conclusion.}} 
Before conclusion, we would like to comment on the azimuthal $\cos(2\phi)$ asymmetry in multiple jets production beyond dijet case. First, let us examine the case for three jet production. Applying a similar equation of~(\ref{eq:sigma2gg}) and the helicity amplitude rule~\cite{Parke:1986gb,Berends:1987me}, it is straightforward to show that $\hat\sigma_2$ vanishes at the tree-level for $gg\to ggg$ too. At one-loop order, it will receive contributions from the interference between ${\cal A}^{(1)}(++++-)$ and ${\cal A}^{(0)}(--++-)$. The nontrivial color structures of these amplitudes lead to a nonzero $\langle \cos(2\phi)\rangle$ for three jet production at $\alpha_s$ order.

For four jet production, however, $\cos(2\phi)$ asymmetry starts at the leading order, where $\hat\sigma_2$ comes from, e.g., ${\cal A}^{(0)}(++++--){\cal A}^{(0)*}(--++--)$ and these are the amplitudes that contribute to $\hat\sigma_0$ too. Therefore. the $\cos(2\phi)$ asymmetry is order 1. The same conclusion holds for five and more jets production. In Table~\ref{table:asym}, we summarize this power counting result of $\cos(2\phi)$ for jet production.
\begin{table}[h]
    \centering
    \def\arraystretch{1.5}
    \begin{tabular}{>{\centering\arraybackslash} m{3.25cm}|>{\centering\arraybackslash} m{1.25cm} >{\centering\arraybackslash} m{1.25cm} >{\centering\arraybackslash} m{1.25cm}}
    \hline\hline
    Number of Jets & 2 & 3 & $\ge$ 4 \\ \hline
    $\langle\cos(2\phi)\rangle$ asymmetry & ${\cal O}(\alpha_s^2)$ &  ${\cal O}(\alpha_s)$  & ${\cal O}(1)$ \\
    \hline \hline
    \end{tabular}
    \caption{\label{table:asym} A brief summary of the power counting analysis for the long range azimuthal angular asymmetries of $\cos(2\phi)$ for jet production in proton-proton collisions.}
  \vspace{-2.ex}
  \end{table}

The $\alpha_s$ hierarchy of $\langle \cos(2\phi)\rangle $ depending on the number of jets is a unique prediction from the NEEC gluon distributions and the helicity structure of parton scattering amplitudes in QCD. It is crucial to test this prediction and pave the way to decipher the underlying mechanism for the nearside ridge in high multiplicity events of $pp$ collisions. In particular, the higher multiplicity events may favor a higher number of jets, where the $\cos(2\phi)$ asymmetries will be larger, in qualitative agreement with the experiment observation~\cite{CMS:2010ifv}. 

In summary, we studied forward-backward azimuthal angular correlations in NEEC measurements in association with multi-jet production in the central rapidity region at the LHC. 
We demonstrated a peculiar prediction for multi-jet production, as a result of double helicity-flip for the $\cos(2\phi)$ asymmetry and the helicity structures of parton scattering amplitudes in QCD, where $\langle \cos(2\phi)\rangle$ starts at ${\cal O}(\alpha_s^2)$ for dijet, ${\cal O}(\alpha_s)$ for three jet, and ${\cal O}(1)$ for four (and more) jet productions. Experimental verification of these predictions will be crucial to establish the NEEC physics at colliders and pave a method to understand the nearside ridge, a long standing puzzle, observed in high multiplicity events in $pp$ collisions at the LHC. 

An important next step is to implement the spinning gluon idea in multi-jet production events in $pp$ collisions. In our calculations, we have applied the energy weighting for the forward and backward hadrons. Two complementary approaches can be applied from our results. First, we can experimentally test the predictions presented in this paper, shown in Figs.~(\ref{fg:dijet2phi}) and (\ref{fg:dijet}). Second, one can apply the existing simulation codes for multi-gluon productions to estimate the $\cos(2\phi)$ asymmetries and compute their contributions to the nearside ridge in $pp$ collisions. This can be either done through a modification of PYTHIA event generator or a combination of numeric simulations from various codes in the literature. Applying a similar idea in the final states~\cite{Chen:2020adz, Chen:2021gdk, Karlberg:2021kwr,Yu:2021zmw} to high multiplicity events in $e^+e^-$ annihilation process~\cite{Chen:2023njr} will be highly desired as well. 

\begin{acknowledgments}


\textbf{\textit{Acknowledgement.}} 
We thank Hua-Xing Zhu for collaboration at the early stage of this project, and Werner Vogelsang for stimulating discussions and comments. We also thank Wei Li for constructive comments and discussions. This work is supported by the Natural Science Foundation of China under contract No.~12175016 (X.~L.), the Office of Science of the U.S. Department of Energy under Contract No. DE-AC02-05CH11231 and under the umbrella of the Quark-Gluon Tomography (QGT) Topical Collaboration with Award DE-SC0023646 (Y.G and F.Y.).  

 \end{acknowledgments}

\bibliographystyle{h-physrev}   
\bibliography{refs}

\newpage    
\appendix

\section{Long Range $\cos(2\phi)$ Asymmetries in the Nucleon EEC associated with Dijet Production at $\alpha_s^2$ Order}

Therefore, we have to go beyond the NLO to derive a nonzero contribution for $\cos(2\phi)$ asymmetry in dijet production. At the NNLO, we have
\begin{eqnarray}
    \hat\sigma_2^{(2)}&\propto &A^{(1)}(+++-)A^{(1)*}(--+-) \nonumber\\
    &+&A^{(1)}(++++)A^{(1)*}(--++)\nonumber\\
    &+&A^{(2)}(++++)A^{(0)*}(--++)+h.c. 
\end{eqnarray}
The first term is easy to derive and the contribution is finite. However, the last two terms contain infrared (IR) divergences. These divergences shall be cancelled by the real soft gluon radiations. To show this more clearly, let us write down the NLO contribution $\hat\sigma_2^{(1)}$ explicitly,
\begin{equation}
    \hat{\sigma}_2^{(1)}\propto  \left<A^{(1)}(++++)|A^{(0)}(--++)\right> + h.c.=\mathcal{O}(\epsilon)\ ,
\end{equation}
which vanishes as $\epsilon\to 0$ with $\epsilon$ being the regulator of dimensional regularization. However, there are linear $\epsilon$ terms in $A^{(1)}(++++)$ which will potentially generate IR divergence with a real soft gluon radiation. Additional contributions come from anomalous soft gluon radiation due to color space modification. These two types of contributions happen to both virtual and real gluon radiations. 

\subsection{Virtual Contributions}

Virtual contributions come from two-loop amplitude of $A^{(2)}(++++)$ and one-loop $A^{(1)}(++--)$. The $\hat \sigma_2^{(2)}$ of $gg\to gg$ will have infrared (IR) divergence at NNLO, even though it vanishes at NLO. There are two contributions to such IR divergence, as will be shown below. To show the IR structure of the NNLO $\hat{\sigma}_2^{(2)}$, recall that the singular behavior of QCD amplitudes has been well established at two loop order, based on the original development in \cite{Catani:1998bh}. Following the notations therein, one expands the amplitude as
\begin{equation}
    \ket{\mathcal{M}(\mu;\{p\})} =4\pi\alpha_S(\mu)\sum_{n=0}\left(\frac{\alpha_S(\mu)}{4\pi}\right)^n \ket{\mathcal{M}^{(n)}(\mu;\{p\}}\ ,
\end{equation}
where the $\ket{\mathcal{M}}$ notation implies that the amplitudes are vectors in color space besides their helicity dependence. More specifically, the amplitude $\mathcal{M}$ can be written in terms of a set of color-space basis as,
\begin{equation}
    \mathcal{M}^{abcd}(\mu;\{p\}) = \sum_{i} \mathcal{C}_i^{abcd} \mathcal{M}_i(\mu;\{p\})\ ,
\end{equation}
such that $\ket{\mathcal{M}}=(\mathcal{M}_1,\mathcal{M}_2,\cdots)$. The amplitude square can be written 
\begin{eqnarray}
    \left<\mathcal{M}|\mathcal{M}\right>&\equiv&   
    \sum_{abcd}\sum_{i,j}  \mathcal{M}_i^*(\mu;\{p\})\mathcal{C}_i^{abcd*}\mathcal{C}_j^{abcd} \mathcal{M}_j(\mu;\{p\})\nonumber \\  
    &=& \sum_{i,j}\mathcal{M}_i^*(\mu;\{p\}) {\cal C\!C}_{ij} \mathcal{M}_j(\mu;\{p\}) \ ,
\end{eqnarray}
where 
\begin{equation}
    {\cal C\!C}_{ij} \equiv \sum_{abcd}\mathcal{C}_i^{abcd*}\mathcal{C}_j^{abcd}\ ,
\end{equation}
is the color space metric tensor.

With these color space notation, the one-loop and two-loop amplitudes can be written as,
\begin{eqnarray}
\ket{\mathcal{M}_{m}^{(1)}(\mu^2;\{p\})}_{\rm{R.S.}} &=& \boldsymbol{I}^{(1)}(\epsilon,\mu;\{p\})\ket{\mathcal{M}_{m}^{(0)}(\mu^2;\{p\})}_{\rm{R.S.}} \nonumber\\
&+&\ket{\mathcal{M}_{m}^{(1)\rm{fin}}(\mu^2;\{p\})}_{\rm{R.S.}}\ ,
\end{eqnarray}
\begin{eqnarray}
\ket{\mathcal{M}_{m}^{(2)}(\mu^2;\{p\})}_{\rm{R.S.}} &=& \boldsymbol{I}^{(1)}(\epsilon,\mu;\{p\})\ket{\mathcal{M}_{m}^{(1)}(\mu^2;\{p\})}_{\rm{R.S.}} \nonumber\\
&+&\boldsymbol{I}^{(2)}(\epsilon,\mu;\{p\})\ket{\mathcal{M}_{m}^{(0)}(\mu^2;\{p\})}_{\rm{R.S.}} \nonumber \\
&+&\ket{\mathcal{M}_{m}^{(2)\rm{fin}}(\mu^2;\{p\})}_{\rm{R.S.}}\ ,
\end{eqnarray}
where all IR divergence can be factorized into the universal structure $\boldsymbol{I}^{(1)}(\epsilon,\mu;\{p\})$ and $\boldsymbol{I}^{(2)}(\epsilon,\mu;\{p\})$. The bold $\boldsymbol{I}^{(1)}$ and $\boldsymbol{I}^{(2)}$ implies that they are generally matrices in color space whose explicit expressions can be found therein. The $\rm{R.S.}$ subscript indicates the scheme dependence of these amplitudes.

Later in \cite{Sterman:2002qn}, it was shown that these IR singularity can be further factorized into color-independent jet function and color-correlated soft function:
\begin{equation}
    \boldsymbol{I}^{(1)}\equiv \Bigg[-\sum_i\left(\frac{\gamma_K^{[i](1)}}{2\epsilon^2}+\frac{\mathcal{G}_0^{[i](1)}}{\epsilon}\right)\boldsymbol{1}+\frac{\boldsymbol{\Gamma}^{(1)}}{\epsilon}\Bigg]\left(-\frac{\mu^2}{s}\right)^\epsilon \ .
\end{equation}
The first term associated with the identity matrix $\boldsymbol{I}$ is color-uncorrelated. It origins from the jet function of each parton in the process and thus sums over all partons $i$. The second term is collected by the so-called soft anomalous dimension matrix $\boldsymbol{\Gamma}^{(1)}$ which is color-correlated. It depends on the process, and can be generally written as~\cite{Aybat:2006wq},
\begin{equation}
    \boldsymbol{\Gamma}^{(1)}=\frac{1}{2}\sum_i \sum_{j\not = i} \boldsymbol{T}_i \cdot \boldsymbol{T}_j \ln\left(\frac{-\mu^2}{s_{ij}}\right)\ ,
\end{equation}
where $s_{ij}=(p_i+p_j)^2$ in all outgoing notation. 

The IR-divergence of the NNLO $\hat{\sigma}$ also origins from such universal IR structure in the amplitudes. Consider that the asymmetry $\hat{\sigma}_2$ can be expanded in the form of, 
\begin{equation}
    \hat{\sigma}_2=(4\pi\alpha_S(\mu))^2 \sum_{n=0} \left(\frac{\alpha_S(\mu)}{4\pi}\right)^n  \hat{\sigma}_2^{(n)}(\mu;\{p\})\ ,
\end{equation}
as well. At NLO, the $\hat{\sigma}_2^{(1)}$ is found to vanish up to $\mathcal{O}(\epsilon)$ terms due to the color space cancelation shown above:
\begin{equation}
    \hat{\sigma}_2^{(1)}= \frac{2}{\mathcal{V}}\left<A^{(1)}(++++)|A^{(0)}(--++)\right> + h.c.=\epsilon \hat{\sigma}_2^{(1)\epsilon}\ ,
\end{equation}
where $\epsilon$ is the regulator of dimensional regularization, and $\mathcal {V}=4 (N^2-1)^2$ that averages over initial gluon colors and helicities. Then, the leading contribution will be from the NNLO $\hat{\sigma}$, which can be written as
\begin{eqnarray}
    \hat\sigma_2^{(2)}&=&\frac{4}{\mathcal{V}}\left<A^{(1)}(++++)|\boldsymbol{I}^{(1)}|A^{(0)}(--++)\right> \nonumber\\
    &+& h.c.+\rm{finite~terms} \ ,
\end{eqnarray}
noting that $A^{(0)}(++++)=A^{(0)}(--+-)=0$. The IR divergence of $\hat\sigma_2^{(2)}$ can be completely identified with that of the $\boldsymbol{I}^{(1)}$ as the amplitudes $A^{(0)}(--++)$ and $A^{(1)}(++++)$ are finite. It can be shown that the IR divergence of NNLO $\hat\sigma_2^{(2)}$ can be written as,
\begin{equation}
    \hat\sigma_2^{(2)} = \frac{1}{\epsilon}\left( \hat{\sigma}_{2,J}^{(2)} + \hat{\sigma}_{2,S}^{(2)} \right) + \rm{finite~terms} \ ,
\end{equation}
where
\begin{eqnarray}
    \hat{\sigma}_{2,S}^{(2)}&= &\frac{4}{\mathcal{V}}\left<A^{(1)}(++++)|\boldsymbol{\Gamma}^{(1)}|A^{(0)}(--++)\right> +h.c. \ ,\nonumber \\
    \hat{\sigma}_{2,J}^{(2)}&= &- \sum_i  \gamma_K^{[i](1)} \hat{\sigma}_2^{(1)\epsilon} = - 8 C_A \hat{\sigma}_2^{(1)\epsilon} \ ,
\end{eqnarray}
where in the second line, we use $\gamma_K^{[g](1)}=2C_A$ and sum over the four gluons. 

In the above expression, the $\hat{\sigma}_{2,S}^{(2)}$ term origins from the soft gluon radiation that change the color structure and thus break the color space cancelation, while the $\hat{\sigma}_{2,J}^{(2)}$ term origins from the incomplete color space cancelation at NLO at $\mathcal{O}(\epsilon)$, which turn into the $1/\epsilon$ divergence when combined with the $1/\epsilon^2$ in $\boldsymbol{I}^{(1)}$. The two terms can be separately evaluated. Utilizing the helicity amplitude in \cite{Ahmed:2019qtg} and the soft anomalous dimension matrix in \cite{Glover:2001af},
\begin{widetext}
\begin{equation}
\boldsymbol{\Gamma}^{(1)}=-2\begin{pmatrix} 
N_c(\tS+\tT) & 0 & 0  & (\tT-\tU) & 0 & (\tS-\tU) \\
0 & N_c(\tS+\tU) & 0  & (\tU-\tT) & (\tS-\tT) & 0 \\
0 & 0 & N_c(\tT+\tU)  & 0 & (\tT-\tS) & (\tU-\tS) \\
2(\tS-\tU) & 2(\tS-\tT) & 0  & 2N_c\tS & 0 & 0 \\
0 & 2(\tU-\tT) & 2(\tU-\tS)  & 0 & 2N_c\tU & 0 \\
2(\tT-\tU) & 0 & 2(\tT-\tS)  & 0 & 0 & 2N_c\tT
\end{pmatrix}\ ,
\end{equation}
where
\begin{equation}
    \tS= \ln\left(\frac{-\mu^2}{\hat{s}}\right)\ ,\tT= \ln\left(\frac{-\mu^2}{\hat{u}}\right)\ , \tU= \ln\left(\frac{-\mu^2}{\hat{t}}\right)\ .
\end{equation}
one has 
\begin{eqnarray}
    \hat{\sigma}_{2,S}^{(2)}&= & \frac{8 N\left(30N+(6+N^2)(N-N_f)\right)}{3(N^2-1)}\left(\frac{\ln{\bar x}}{x}+\frac{\ln{x}}{\bar x}\right)\ , \\
    \hat{\sigma}_{2,J}^{(2)}&= &-8 N\hat{\sigma}_2^{(1)\epsilon}= (-8N)\frac{N^2(N_f-N)}{3(N^2-1) }\left(\pi^2-2\ln\bar x\ln x+\frac{1+x^2}{x^2}\ln^2{\bar x}+\frac{1+\bar x^2}{{\bar x}^2}\ln^2x+ \frac{4\ln\bar x}{x}+\frac{4\ln x}{\bar x}  \right)\ ,
\end{eqnarray}
where $x=-\hat{t}/\hat{s}$ and $\bar x = 1-x$.
\end{widetext}
We note that the above $6\times 6$ matrix is written with respect to basis:
\begin{eqnarray}
    {\cal C}_1&=&{\rm tr}\left[T^aT^bT^cT^d\right]+{\rm tr}\left[T^aT^dT^cT^b\right] \ , \\
    {\cal C}_2&=&{\rm tr}\left[T^aT^bT^dT^c\right]+{\rm tr}\left[T^aT^cT^dT^b\right] \ , \\
    {\cal C}_3&=&{\rm tr}\left[T^aT^cT^bT^d\right]+{\rm tr}\left[T^aT^dT^bT^c\right] \ ,\\
    {\cal C}_4&=&{\rm tr}\left[T^aT^b\right]{\rm tr}\left[T^cT^d\right] \ ,\\
    {\cal C}_5&=&{\rm tr}\left[T^aT^c\right]{\rm tr}\left[T^bT^d\right] \ ,\\
    {\cal C}_6&=&{\rm tr}\left[T^aT^d\right]{\rm tr}\left[T^bT^c\right] 
    \ .
\end{eqnarray}
Such IR divergence must be canceled by the real soft radiation, as will be shown below.

\subsection{Real Contributions}

To derive the real soft gluon contributions, we write down explicitly the amplitudes in the following manner,
\begin{eqnarray}
    A^{(1)}(++++)&=&{\cal F}_1{\cal C}_1+{\cal F}_2{\cal C}_2+{\cal F}_3{\cal C}_3\nonumber \\
    &&+{\cal F}_4{\cal C}_4+{\cal F}_5{\cal C}_5+{\cal F}_6{\cal C}_6 \ ,\\
    A^{(0)}(++--)&=&{\cal G}_1{\cal C}_1+{\cal G}_2{\cal C}_2+{\cal G}_3{\cal C}_3 \ .
\end{eqnarray}
In the above expansion, ${\cal G}_i$ do not contain $\epsilon$,
\begin{eqnarray}
    {\cal G}_1=\frac{4}{1-x}\ ,
   ~~ {\cal G}_2=\frac{4}{x}\ ,
    ~~{\cal G}_3=-\frac{4}{x(1-x)}\ ,
\end{eqnarray}
while ${\cal F}_i$ do have $\epsilon$-expansion terms,
\begin{eqnarray}
    {\cal F}_{1,2,3}&=&\frac{4}{3}(N_f-N)+\epsilon {\cal F}_{1,2,3}^\epsilon (x)+\epsilon^2 {\cal F}_{1,2,3}^{\epsilon^2} (x)\ ,\\
    {\cal F}_{4,5,6}&=&-8+\epsilon {\cal F}_{4,5,6}^\epsilon (x)+\epsilon^2 {\cal F}_{4,5,6}^{\epsilon^2} (x)\ \ .
\end{eqnarray}
With the above results, we can clearly check that $\langle A^{(1)}(++++)|A^{(0)}(++--)\rangle$ is proportional to ${\cal G}_1+{\cal G}_2+{\cal G}_3$ and vanishes at $\epsilon=0$. However, because ${\cal F}_i^{\epsilon}$ are different, it does not vanish at ${\cal O}(\epsilon)$ as discussed above.  

Real soft gluon radiation can derived following that in Ref.~\cite{Sun:2015doa}. It can attach to all external gluons. By choosing a particular gauge for the soft gluon, $\epsilon(k_g)\cdot p_2=0$, we can eliminate the gluon attachment to $p_2$ line. Therefore, the soft gluon radiation amplitude can be written as,
One soft gluon radiation takes the form,
\begin{eqnarray}
{\cal S}^\mu_g([\cdots]^{abcd})&\equiv&\frac{2k_1^\mu}{2k_1\cdot k_g}f_{gcf}\left[\cdots \right]^{abfd} \nonumber\\
&+&\frac{2k_2^\mu}{2k_2\cdot k_g}f_{gdf}\left[\cdots \right]^{abcf}  \nonumber\\
&+&\frac{2p_1^\mu}{2p_1\cdot k_g}f_{gaf}\left[\cdots \right]^{fbcd}  \ ,
\end{eqnarray}
where $\left[\cdots \right]^{abcd}$ represents the amplitude $A^{(1)}(++++)$ and $A^{(0)}(++--)$ with associated color indices and $g$ and $\mu$ represents the color and Lorentz indices of the emitted soft gluon. From that,
we can derive the soft gluon radiation contribution easily, with the polarization
tensor for the radiated gluon,
\begin{equation}
\Gamma^{\mu\nu}(k_g)=\left(-g^{\mu\nu}+\frac{k_g^\mu p_2^\nu+k_g^\nu p_2^\mu}{k_g\cdot p_2}\right)  \ .
\end{equation}
For example, from the amplitude squared of the soft gluon radiation
terms in the above, we have
\begin{eqnarray}
\frac{2p_1^\mu}{2p_1\cdot k_g}\frac{2p_1^\nu}{2p_1\cdot k_g}\Gamma_{\mu\nu}
&=&S_g(p_1,p_2)\ ,\\
\frac{2k_1^\mu}{2k_1\cdot k_g}\frac{2k_1^\nu}{2k_1\cdot k_g}\Gamma_{\mu\nu}
&=&S_g(k_1,p_2)\ ,\\
\frac{2k_2^\mu}{2k_2\cdot k_g}\frac{2k_2^\nu}{2k_2\cdot k_g}
\Gamma_{\mu\nu}
&=&S_g(k_2,p_2)\ ,
\end{eqnarray}
where $S_g(p,q)$ is a short-handed notation for
\begin{equation}
S_g(p,q)=\frac{2p\cdot q}{p\cdot k_gq\cdot k_g} \ .
\end{equation}
Similarly, we derive the interferences between them,
\begin{eqnarray}
2\frac{2k_1^\mu}{2k_1\cdot k_g}\frac{2p_1^\nu}{2p_1\cdot k_g}\Gamma_{\mu\nu}
&=&S_g(k_1, p_2)+S_g(p_1, p_2)-S_g(k_1, p_1)\ ,\nn\\
2\frac{2k_2^\mu}{2k_2\cdot k_g}\frac{2p_1^\nu}{2p_1\cdot k_g}\Gamma_{\mu\nu}
&=&S_g(k_2, p_2)+S_g(p_1, p_2)-S_g(k_2, p_1)\ ,\nn\\
2\frac{2k_1^\mu}{2k_1\cdot k_g}\frac{2k_2^\nu}{2k_2\cdot k_g}\Gamma_{\mu\nu}
&=&S_g(k_1, p_2)+S_g(k_2, p_2)-S_g(k_1, k_2)\ .\nn
\end{eqnarray}
To evaluate these soft gluon radiation contributions at small transverse momentum $q_\perp$, we integrate out the phase space of longitudinal momentum fraction $x_g$. This integral normally generates large logs, e.g.,
\begin{equation}
    \int_{x_{min}}\frac{dx_g}{x_g} \Rightarrow \ln\frac{\hat s}{q_\perp^2}\ ,
\end{equation}
where $x_{min}\sim q_\perp^2/\hat s$ is the lower limit from the kinematic constraint of the real gluon radiation. Performing these phase space integral, we will obtain the following contributions,
\begin{eqnarray}
S_g(p_1,p_2)  &\Rightarrow& \frac{1}{q_\perp^2}\left[2 \ln\frac{\hat s}{q_\perp^2}\right],  \nn\\ 
S_g(k_1,p_1) &\Rightarrow& \frac{1}{q_\perp^2}\left[\ln\frac{\hat s}{q_\perp^2}+\ln\frac{\hat t}{\hat u}+\ln\frac{1}{R_1^2}+\frac{\epsilon}{2}\ln^2\frac{1}{R_1^2}\right],  \nn\\
S_g(k_2,p_1) &\Rightarrow& \frac{1}{q_\perp^2}\left[\ln\frac{\hat s}{q_\perp^2}+\ln\frac{\hat u}{\hat t}+\ln\frac{1}{R_2^2}+\frac{\epsilon}{2}\ln^2\frac{1}{R_2^2}\right], \nn\\
S_g(k_1,p_2) &\Rightarrow& \frac{1}{q_\perp^2}\left[\ln\frac{\hat s}{q_\perp^2}+\ln\frac{\hat u}{\hat t}+\ln\frac{1}{R_1^2}+\frac{\epsilon}{2}\ln^2\frac{1}{R_1^2}\right], \nn\\
S_g(k_2,p_2) &\Rightarrow& \frac{1}{q_\perp^2}\left[\ln\frac{\hat s}{q_\perp^2}+\ln\frac{\hat t}{\hat u}+\ln\frac{1}{R_2^2}+\frac{\epsilon}{2}\ln^2\frac{1}{R_2^2}\right],\label{dis} \nn\\
S_g(k_1,k_2) &\Rightarrow& \frac{1}{q_\perp^2}\left[2\ln\frac{\hat s^2}{\hat t\hat u}+\ln\frac{1}{R_1^2R_2^2}+\frac{\epsilon}{2}\left(\ln^2\frac{1}{R_2^2}\right.\right.\nn\\
&&\left.
\left.+\ln^2\frac{1}{R_2^2}-8\ln\frac{\hat s}{-\hat t}\ln\frac{\hat s}{-\hat u}\right)\right], 
\end{eqnarray}
where a narrow jet approximation has been applied to derive the jet contribution with $R_1$ and $R_2$ to regulate the jet divergences associated with $k_1$ and $k_2$ for the final state particles, respectively. 

Then the soft radiation contribution can be generally written as:
\begin{eqnarray}
    & &\sum_{a,b,c,d,g} {\cal S}_g^{\mu*}([\cdots]^{abcd}) {\cal S}_g^{\nu}([\cdots]^{abcd}) \Gamma_{\mu\nu} \nonumber\\
    &=&  \sum_{a,b,c,d}[\cdots]^{abcd*} {\cal S\!\!S}_{abcd} [\cdots]^{abcd}
\end{eqnarray}
we define the color-space matrix for the soft radiation $\overline{\cal S\!\!S}$ such that,
\begin{equation}
    \left<[\cdots]|\overline{\cal S\!\!S}|[\cdots]\right>= \sum_{a,b,c,d}[\cdots]^{abcd*} {\cal S\!\!S}_{abcd} [\cdots]^{abcd}\ .
\end{equation}
The soft radiation matrix $\overline{\cal S\!\!S}$ carries all structure of the soft radiation, independent of the amplitude $[\cdots]^{abcd}$. Following the notation in the previous subsection, one has
\begin{equation}
    \sum_{k}{\cal C\!\!C}_{ik}  \overline{{\cal S\!\!S}}_{kj} =\sum_{a,b,c,d}\text{Tr}[{\cal C}_i^{abcd*}] {\cal S\!\!S}^{abcd}\text{Tr}[{\cal C}^{abcd}_j]\ .
\end{equation}
Then one find that the $\overline{{\cal S\!\!S}}_{jk}$ has a simple expression in terms of all the $S_g(p,q)$s:
\begin{eqnarray}
     \overline{{\cal S\!\!S}}&=&\frac{ 4\pi \epsilon}{ q_\perp^2}\Big[4N \left(\ln^2(R_1)+\ln^2(R_2)\right)\boldsymbol{1}+\ln\frac{\hat s}{-\hat t}\ln\frac{\hat s}{-\hat u} \boldsymbol{L} \Big] \nonumber \\
     &+&\frac{4\pi}{ q_\perp^2}\left[2\text{Re}[\boldsymbol{\Gamma}^{(1)}] +4N_c \ln\frac{\hat s \mu^4}{q_\perp^2 R_1R_2 \hat t\hat u}\boldsymbol{1}\right]\ ,
\end{eqnarray}
where the $\mu$-dependence gets canceled combining the two terms in the second line. Explicitly contraction of the above matrix with the helicity amplitudes leads to
\begin{eqnarray}
    &&    \frac{\alpha_s}{2\pi^2}\frac{1}{q_\perp^2}\frac{4}{{\cal V}}\left\{
    -{\cal N}_0 
    \left[{\cal G}_1\ln\frac{1}{x}+{\cal G}_2\ln\frac{1}{1-x}\right]\right.\nn\\
    &&   + \epsilon \left[N{\cal M}_0^\epsilon\ln\frac{\hat s^2}{q_\perp^4R_1^2R_2^2}+ 2N\left({\cal M}_1^\epsilon \ln\frac{1}{x}+{\cal M}_2^\epsilon \ln\frac{1}{1-x}\right)\right.\nn\\
    &&\left.\left.+{\cal N}_0\left({\cal G}_1+{\cal G}_2\right)\ln\frac{1}{x}\ln\frac{1}{1-x}\right] +
    \epsilon^2N{\cal M}_0^{\epsilon^2}\ln\frac{\hat s^2}{q_\perp^4}+\cdots\right\} \nn\ ,\\ \label{eq:e112}
\end{eqnarray}
where ${\cal N}_0=(N^2-1)(N^2+6)(N^2-N_f)$ and 
\begin{eqnarray}
    {\cal M}_0^\epsilon &=&N\left[3 ({\cal F}_1^\epsilon{\cal G}_1+{\cal F}_2^\epsilon{\cal G}_2+{\cal F}_3^\epsilon{\cal G}_3)+({\cal F}_4^\epsilon-{\cal F}_5^\epsilon) {\cal G}_1\right.\nn\\
    &&\left. +({\cal F}_4^\epsilon-{\cal F}_6^\epsilon){\cal G}_2)\right] \ ,\\
    {\cal M}_0^{\epsilon^2} &=&N\left[3 ({\cal F}_1^{\epsilon^2}{\cal G}_1+{\cal F}_2^{\epsilon^2}{\cal G}_2+{\cal F}_3^{\epsilon^2}{\cal G}_3)+({\cal F}_4^{\epsilon^2}-{\cal F}_5^{\epsilon^2}) {\cal G}_1\right.\nn\\
    &&\left. +({\cal F}_4^{\epsilon^2}-{\cal F}_6^{\epsilon^2}){\cal G}_2)\right] \ ,\\
    {\cal M}_1^\epsilon &=& {\cal G}_1\left(11 {\cal F}_1^\epsilon+2{\cal F}_2^\epsilon+2{\cal F}_3^\epsilon+6{\cal F}_4^\epsilon+3{\cal F}_5^\epsilon+6{\cal F}_6^\epsilon\right),\\
    {\cal M}_2^\epsilon &=& {\cal G}_2\left(2 {\cal F}_1^\epsilon+11{\cal F}_2^\epsilon+2{\cal F}_3^\epsilon+6{\cal F}_4^\epsilon+6{\cal F}_5^\epsilon+3{\cal F}_6^\epsilon\right),
   \end{eqnarray}
where we have applied ${\cal G}_1+{\cal G}_2+{\cal G}_3=0$ to simplify the above expressions.

Meanwhile, there are collinear radiation contributions from the two incoming gluons. At low $q_\perp\ll P_\perp$, they can be written as 
\begin{eqnarray}
    &&    \frac{\alpha_s}{2\pi^2}\frac{1}{q_\perp^2}\frac{4}{{\cal V}}\epsilon 2N{\cal M}_0^\epsilon\int \frac{d\xi}{\xi}\frac{d\xi'}{\xi'}x_ax_b d_{g,\rm EEC}(x_a') d_{g,\rm EEC}(x_b')\nn\\
    &&\times \left[ {\cal P}^d_{gg}(\xi)\delta(1-\xi')+{\cal P}^d_{gg}(\xi')\delta(1-\xi) \right] \ ,
\end{eqnarray}
where $x_a'=x_a/\xi$ and $x_b'=x_b/\xi'$ and ${\cal P}_{gg}^d(\xi)=\xi/(1-\xi)_++\beta_0\delta(1-\xi)$. 

All the real gluon radiations will lead to IR divergences when we integrate over $q_\perp$,
\begin{eqnarray}
&\!\!&\!\!\int_{q_\perp<q_0}\frac{d^{D-2}q_\perp}{(2\pi)^{D-2}}\frac{1}{q_\perp^2}=\frac{1}{4\pi}\left[-\frac{1}{\epsilon}+\ln\frac{q_0^2}{\mu^2}\right ]\ ,\\
&\!\!&\!\!\int_{q_\perp<q_0}\frac{d^{D-2}q_\perp}{(2\pi)^{D-2}}\frac{1}{q_\perp^2}\ln\frac{\hat s}{q_\perp^2}=\frac{1}{4\pi}\left[\frac{1}{\epsilon^2}-\frac{1}{\epsilon}\ln\frac{\hat s}{\mu^2}\right.\nn\\
&&\left.
+\frac{1}{2}\left(\ln\frac{\hat s}{\mu^2}\right)^2-\frac{1}{2}\left(\ln\frac{\hat s}{q_0^2}\right)^2-\frac{\pi^2}{12}\right ] \ .
\end{eqnarray}
Substituting the above integrals into the real gluon radiation contributions, we will obtain the IR divergences plus the respective finite terms. The IR divergences will be completely cancelled by the relevant virtual contributions. It is interested to note that some of finite terms also completely cancel out each between the real and virtual contributions, for example, those associated with ${\cal M}_{1,2}^\epsilon$, ${\cal M}_0^{\epsilon^2}$ and part of ${\cal M}_0^{\epsilon}$. The final finite terms contain three important contributions: one from the soft anomalous dimension matrix $\Gamma^{(1)}$, one from the collinear splitting, and the last one from the finite terms in the virtual contributions. These contributions are summarized in Eq.~(\ref{eq:dijetfinal}), where $\sigma_{2,2}^{(2)}$ can be further decomposed as,
    \begin{equation}
    \hat{\sigma}_{2,2}^{(2)} 
    \equiv 
    -\frac{8\mathcal{N}_1  N\ln x \ln \bar x}{3(N^2-1)x\bar x}+\hat{\sigma}_{2,2,1}^{(2)}+\hat{\sigma}_{2,2,2}^{(2)}+\hat{\sigma}_{2,2,3}^{(2)} \ ,
\end{equation}
where $\mathcal{N}_1\equiv (N^2+36)N-(N^2+6)N_f$. In the above equation, $\hat\sigma_{2,2,1}^{(2)}$ comes from the finite term in the amplitude interference of $A^{(1)}(+++-)A^{(1)}(--+-)$, $\hat\sigma_{2,2,2}^{(2)}$ from $A^{(1)}(++++)A^{(1)}(--++)$, and $\hat\sigma_{2,2,3}^{(2)}$ from $A^{(2)}(++++)A^{(0)}(--++)$. The explicit expressions for those terms are 
\begin{widetext}
\begin{align}
    \hat{\sigma}_{2,2,1}^{(2)} &=\frac{1}{9(N^2-1)}\left[4N^2(N-N_f)^2\left(\frac{x}{\bar x^2}+\frac{\bar x}{x^2}-x\bar x\right)+18\left(N^2(N^2+12)-2N(N^2+2)N_f+(N^2+\frac{6}{N^2}-2)N_f^2\right)\right] \ ,\\
    \hat{\sigma}_{2,2,2}^{(2)} &= \frac{-1}{N^2(N^2-1)}\left[N^3\mathcal{N}_1\left(
    \frac{4\ln{x}\ln{\bar x}}{3x\bar x}+\pi^2+\ln^2\frac{\bar x}{x}\right)+\frac{\mathcal{N}_2}{3} \left(2+2\ln\frac{\bar x}{x}(x-\bar x)+(1-2x\bar x) \left(\pi^2+\ln^2\frac{x}{\bar x}\right)\right)\right] \ , \\ 
    \hat{\sigma}_{2,2,3}^{(2)} &=\frac{1}{9(N^2-1)}\Big[4N^2(N-N_f)^2\left(1-\frac{1}{x\bar x}\right)^2-\frac{6 N \mathcal{N}_1} {x\bar x}(\pi^2+x\ln^2 x+\bar x \ln^2 \bar x)\Big]\ ,
\end{align}
where $\mathcal{N}_2\equiv (N^4-6N^2+18)N_f^2+(N^2+36)N^4-2N^3 (N^2+6)N_f$.
\end{widetext}

\end{document}